\documentclass[12pt,a4paper]{article}
\usepackage[english]{babel}
\usepackage{setspace}
\usepackage{rotating}

\begin{document}

\doublespacing
\begin{titlepage}
\title{{\bf Liquid-vapor coexistence in square-well fluids: an RHNC study }}

\author{ {\bf Achille Giacometti}$^{\rm a}$, {\bf Giorgio Pastore}$^{\rm b*}$, and
{\bf Fred Lado}$^{\rm c}$}

\date{\today}
\maketitle

\begin{abstract}
We investigate the ability of the reference hypernetted-chain integral equation
to describe the phase diagram of square-well fluids with four different 
ranges of attraction. Comparison of our results with simulation data 
shows that the theory is able to reproduce with fairly good accuracy a 
significant part of the coexistence curve, 
provided an extrapolation procedure is used to circumvent the well-known 
pathologies of the pseudo-spinodal line, which are more severe at 
reduced width of the attractive well. The method provides a useful 
approach for a quick assessment of the location of the liquid-vapor 
coexistence curve in this kind of fluid and serves as a check for the 
more complex problem of anisotropic ``patchy'' square-well molecules.

\end{abstract}

Pacs numbers: 64.70.F-,  61.20.Gy, 64.60.Ej, 65.20.Jk


\vfill
 
\noindent $^{\rm a}$Dipartimento di Chimica Fisica, Universit\`a Ca' Foscari, S. Marta DD 2137, 
I-30123 Venezia, Italy

\noindent $^{\rm b}$Dipartimento di Fisica Teorica dell' Universit\`a,
Strada Costiera 11, 34014 Trieste, Italy
 
\noindent $^{\rm c}$Physics Department, North Carolina State University, Raleigh, North Carolina 
27695-8202

\noindent $^*$Corresponding author. Email: pastore@ts.infn.it 
 
\end{titlepage}
\section*{Introduction}
Structural and thermodynamic properties of the square-well (SW) fluid 
have been studied with a huge variety of statistical mechanical methods. 
This system has long been viewed as the simplest nontrivial model able 
to capture the main phenomenology of real atomic fluids by complementing 
hard-sphere repulsion with a finite-range and constant attractive well \cite{BH}.  
Almost all theories for the liquid state have been 
applied to the study of the SW fluid and many numerical simulation 
studies have been published  
\cite{Henderson,Vega,deMiguel,delRio,Kim-Fisher,Singh,Largo,Liu,Pagan,MDKiselev,
Lopez,White,Voertler}, transforming this system into an important 
testbed for theories 
\cite{Henderson,Smith-Henderson,Gil-Villegas,Caccamo,RG,Kahl,
Scholl-Paschinger}.

More recently, the study of colloids and protein solutions has 
renewed interest in SW fluids 
as a tool  to study trends and general features of strongly 
localized isotropic attractions.
However, more detailed investigations of such systems suggest that the 
isotropic model should be modified to account for  highly directional 
(patchy) interactions \cite{Sciortino}. 
The recent model studied by Kern and Frenkel \cite{Kern} can be described as 
a SW model where the potential well has finite {\em angular} extent as well 
as finite {\em radial} extent.

A key problem with such patchy fluids is the determination of their 
phase diagram. Specialized computer simulation methods  have been developed to 
allow efficient and reliable determination of liquid-vapor coexistence. While applications to atomic systems is nowadays an almost trivial exercise, numerical simulation in 
the presence of highly attractive and directional forces is still a 
challenging and time-consuming task. Theoretical modeling, albeit 
approximate, may provide a worthwhile tool for faster and wide-ranging scans 
of the relevant parameter space. For such purposes, perturbation methods have 
been used extensively in the past, but their accuracy in the presence of strongly directional attractions is not uniform. Alternatively, modern integral equation theories are known 
to provide reliable and accurate description of structural properties for isotropic \cite{Caccamo,RHNC} as well as anisotropic potentials \cite{Lado_anisotrop} . It is true that 
in the vicinity of a liquid-vapor critical point they manifest shortcomings 
ranging from branching of multiple unphysical solutions \cite{Belloni} 
to wrong critical behavior \cite{Caccamo}. Notwithstanding such limitations, 
we believe that integral equation approach is still able to provide
a quick and reasonably  reliable description of the phase diagram. 
Although results for fluid-fluid coexistence from modern 
integral equation theories are relatively scarce, the existing evidence 
shows that they may provide an accurate description of the low temperature 
part of the binodal curve and, by extrapolation, make it possible to  
approximately locate the critical point in one-component 
\cite{Caccamo,Lomba} and two-component systems \cite{Pastore}. 
More accurate approximations, such as SCOZA or HRT, have been applied to 
the SW fluid \cite{Kahl,Scholl-Paschinger}, but their implementation for 
anisotropic interactions is not an easy task.

Motivated by an interest in applying the reference hypernetted chain (RHNC) 
approximation \cite{RHNC} to simple {\em anisotropic} models of 
patchy colloids, as a preliminary step we wanted to understand the 
limits and shortcomings of the integral equation approach to such systems. 
For this reason, we have undertaken here the study of the liquid-vapor 
coexistence of an {\em isotropic} SW fluid, the extreme limit of patchy SW 
potentials, using the RHNC approximation.

The RHNC approximation has been used before to study SW fluids 
\cite{Gil-Villegas} 
but, so far as we know, it has not been used to study phase coexistence 
in such systems nor it has been tested at temperatures below the critical 
temperature. In this communication, we report results for structural 
properties and liquid-vapor coexistence of a few SW fluids of varying 
attraction range. The plan of the paper is the following:  In section 2, 
we summarize the basic RHNC theory and discuss the computational details of 
our calculation of the phase coexistence curve. Results are presented and 
critically discussed in section 3. A short summary of our findings and 
conclusions is assembled in the final section.

\section{RHNC theory and phase coexistence calculations}

We consider a system of spherical particles of diameter $\sigma $ 
interacting via a pair-wise square-well potential $ \phi(r)$ given by
$$ 
\phi(r) = \left\{ \begin{array}{rl}
           \infty,    & \mbox{ $r < \sigma $} \\
          -\epsilon,  & \mbox{ $\sigma \le r \le \lambda \sigma $}, \\
           0,         & \mbox{ $\lambda \sigma < r $ }
           \end{array} \right.
$$
where $\epsilon$ is the depth and $\lambda$ the dimensionless extent of the 
potential well.
Reduced temperature and density are introduced as usual 
using $\epsilon$ and $\sigma$
as energy and length scale respectively: $T^*=k_B T/\epsilon$, 
$\rho^*=\rho \sigma^3$.

\subsection{RHNC integral equation}

The pair distribution function $g(r)$ of a classical fluid is related to 
the pair potential $\phi(r)$ by the exact relation \cite{Hansen}
\begin{equation}
g(r) = \exp \left[ - \beta \phi(r) + h(r) - c(r) + B(r) \right], 
\label{exact1}
\end{equation}
where $\beta=1/k_B T$ is the inverse Kelvin temperature, $h(r) = g(r)-1$ 
is the pair correlation function, and $c(r)$ is the direct correlation 
function defined via the Ornstein-Zernike (OZ) equation,
\begin{equation}
h(r) = c(r) + \rho \int {\rm d} {\bf r'} c( | {\bf r} - {\bf r'} |) h(r').  
\label{OZ1}
\end{equation}
It is more convenient for numerical work to solve equations 
(\ref{exact1}) and (\ref{OZ1}) for the indirect correlation 
function $\gamma(r)=h(r)-c(r)$ with the OZ equation de-convoluted in 
Fourier space; the pair of equations to solve then becomes
\begin{equation}
c(r) = \exp \left[ - \beta \phi(r) + \gamma(r) + B(r) \right] - 1 - \gamma(r), 
\label{exact2}
\end{equation}
\begin{equation}
\tilde{\gamma}(k) = \frac{\rho \tilde{c}^2(k)}{1-\rho \tilde{c}(k)}.
\label{OZ2}
\end{equation}
These equations are connected through the transforms
\begin{equation}
\tilde{c}(k) = \frac{4\pi}{k} \int_0^\infty dr\, r c(r) \sin (kr)
\label{FTc}
\end{equation}
following equation (\ref{exact2}) and 
\begin{equation}
\gamma(r) = \frac{1}{2\pi^2r} \int_0^\infty dk\, k \tilde{\gamma}(k) \sin (kr)
\end{equation}
following equation (\ref{OZ2}) to form an iteration cycle. The so-called bridge function $B(r)$ that appears above is a complicated functional of the 
pair correlation function for which no exact computable expression is known. 
The various integral equation formulations found in the literature 
correspond to different approximate forms for $B(r)$.

The SW direct correlation function $c(r)$ is of course discontinuous at 
$r=\sigma$ and $r=\lambda\sigma$. To compute its Fourier transform (\ref{FTc}), we assign the single value of $c(r)$ at a discontinuity to be the 
arithmetic mean of its separate values at the discontinuity \cite{Arfken}; e.g.,
\begin{equation}
c(\lambda\sigma) = \lim_{\varepsilon \rightarrow 0} \frac{c(\lambda\sigma+\varepsilon)+c(\lambda\sigma-\varepsilon)}{2}.
\end{equation}
We further ensure that the discontinuities fall on calculated grid points 
in $r$. Moreover, note that it is only the short-ranged direct correlation function 
$c(r)$ that is expected to vanish at large $r$; i.e., its Fourier transform 
requires $c(r_{\rm max}) = 0$, where $r_{\rm max}$ is the cutoff distance 
in the $r$ grid. In the discrete notation used below, $r_{\rm max}=N_r \Delta r$. The pair correlation function $h(r)$, on the other hand, may have whatever long-range tail it wishes without affecting the calculation.

The RHNC closure \cite{RHNC} assumes that the unknown bridge function 
$B(r)$ can be approximated by the corresponding known bridge function 
$B_0(r)$ of a reference system.
Within this approach, just as in the original hypernetted-chain 
closure \cite{M&H}, the excess Helmholtz free energy per particle 
can be written in closed form as \cite{RHNC} 
\begin{eqnarray}
\label{free_energy}
\frac{\beta A_{\rm ex}}{N} &=& 
\frac{\beta A_{1}}{N} +\frac{\beta A_{2}}{N}+\frac{\beta A_{3}}{N},
\end{eqnarray}
where
\begin{eqnarray}
\label{free_energy1}
\frac{\beta A_1}{N} &=& -\frac{1}{2} \rho \int d \mathbf{r} \left\{ \frac{1}{2} h^2(r)+h(r)-g(r) \ln \left [ g(r) e^{\beta \phi(r)} \right] \right\} , \\
\label{free_energy2}
\frac{\beta A_2}{N}&=& - \frac{1}{2 \rho} \int \frac{d\mathbf{k}}{\left(2\pi\right)^3}
\left\{ \ln \left[ 1 + \rho \tilde{h}(k) \right] - \rho \tilde{h}(k) \right\}, \\
\label{free_energy3}
\frac{\beta A_3}{N}&=& \frac{\beta A_3^0}{N}-\frac{1}{2} \rho \int d \mathbf{r} \left[ g(r)-g_0(r) \right] B_0(r).
\end{eqnarray}
Equation (\ref{free_energy3}) directly expresses the RHNC approximation; 
here $A_3^0$ is the corresponding reference system contribution, computed 
from the known free energy $A^0$ of the reference system as 
$A_3^0=A^0-A_1^0-A_2^0$, with $A_1^0$ and $A_2^0$ calculated as above 
but with reference system quantities. 

Length and energy parameters of the reference system, respectively 
$\sigma_0$ and $\epsilon_0$, should be optimized by variational minimization 
to obtain  
the free energy from equation (\ref{free_energy}), 
which leads \cite{Lado} to the conditions (written here in dimensionless forms)
\begin{eqnarray}
\label{ChooseSigma}
\rho \int d\mathbf{r} [g(r) - g_0(r)] \sigma_0 \frac{\partial B_0(r)}{\partial \sigma_0}&=&0, \\
\rho \int d\mathbf{r} [g(r) - g_0(r)] \epsilon_0 \frac{\partial B_0(r)}{\partial \epsilon_0}&=&0.
\end{eqnarray}
These guarantee thermodynamic consistency between the direct calculations of the pressure $P$ and internal energy $U$ using $g(r)$,
\begin{equation}
\label{pressure}
\beta P = \rho - \frac{1}{6}\, \rho^2 \int {\rm d} {\bf r}\,  g(r) r \frac{ d \beta \phi(r)}{dr},  \end{equation}
\begin{equation}
\label{energy}
\frac{U}{N} =  \frac{1}{2}\, \rho \int {\rm d} {\bf r}\,  g(r) \phi(r),  
\end{equation}
and the corresponding thermodynamic derivatives of the free energy, 
$\beta P - \rho = \rho^2 \partial(\beta A_{\rm ex}/N)/\partial\rho$ and 
$U/N = \partial(\beta A_{\rm ex}/N)/\partial\beta$. At present, however, 
only the hard sphere (HS) system, with sphere diameter $\sigma_0$ but no 
energy scale, is well enough known to serve as reference system. With this 
choice, only equation (\ref{ChooseSigma}) is implemented here for the 
results of the next section. 

We will thus obtain the two main ingredients needed, pressure and chemical 
potential, in a consistent way with no additional approximations beyond the 
choice of hard spheres as reference system. Any inaccuracy in the 
computed results can be ascribed solely to the quality of the chosen bridge 
function.

We solve the RHNC equations numerically on $r$ and $k$ grids of 
$N_r=2048$ points with intervals $\Delta r = 0.01 \sigma$ and 
$\Delta k = \pi/(N_r \Delta r)$ using the combination of Newton-Raphson 
and Picard methods introduced by Gillan \cite{Gillan} and optimized by 
Lab\'{i}k et al. \cite{LMV}. Fourier transforms are evaluated with the 
Fast Fourier Transform algorithm. Sample calculations with the same 
$\Delta r = 0.01 \sigma$ but increased $N_r=4096$ and $N_r=8192$ 
produce {\em identical} results for all the printed output; i.e., to four 
or five significant figures. 

For the bridge function, we use the Verlet-\-Weis-\-Henderson-\-Grundke \\
(VWHG) parametrization of HS numerical simulation results \cite{VW,HG}. 
The HS free energy is taken from the Carnahan-Starling equation of state 
\cite{CS}, which is also incorporated into the VWHG parametrization. 
Other parametrizations exist and in the only previous investigation of the 
SW fluid with RHNC, at supercritical temperatures, Gil-Villegas et al. \cite{Gil-Villegas} used that of Malijevsk\'{y} and Lab\'{i}k \cite{ML1,ML2} and found excellent agreement 
with  computer simulation results for thermodynamics and correlation 
functions.  In the present study we choose VWHG  bridge functions 
because they have been frequently used in  RHNC calculations (also for 
liquid-vapor coexistence of HS Yukawa \cite{LA} fluids), thus allowing a direct comparison of the performances of the approximation with different potentials.

\subsection{Liquid-vapor coexistence}

Two-phase coexistence at constant temperature requires the equality of pressure and chemical potential of the two phases; RHNC provides convenient expressions for both quantities,
\begin{equation}
\beta P = \rho + \frac{2}{3}\, \pi \rho^2 \sigma^3 \left[ y(\sigma)e^{\beta\epsilon} - \lambda^3 y(\lambda\sigma)(e^{\beta\epsilon} - 1) \right],  
\end{equation}
\begin{equation}
\beta \mu = \frac{\beta A_{\rm ex}}{N} + \left( \frac{\beta P}{\rho} - 1
\right) + \ln{( \rho \sigma^3)},
\end{equation}
where we have now specialized the pressure equation (\ref{pressure}) for the SW potential and we are using the cavity function $y(r) = g(r)\exp[\beta \phi(r)]$, which has no discontinuities. An additional density-independent term from the ideal gas limit has been neglected in the total chemical potential $\mu$.

In order to find the densities of the two coexisting phases, we use the following 
procedure \cite{Schlijper}: we start two calculations, one at low density and the other at high density at the temperature of interest. Then, density is increased from the lowest value and decreased from the highest until either the coexistence conditions
are satisfied or the numerical program is unable to converge, 
even after 
systematic reduction of the density change,
signaling the disappearance of a solution of the integral equation. 
Conditions of equal pressure and chemical potential are verified by looking for numerical solutions of the nonlinear set of equations,
\begin{eqnarray}
\beta P(\rho_{\rm l}) &=& \beta P(\rho_{\rm g}), \\
\beta \mu(\rho_{\rm l}) &=& \beta \mu(\rho_{\rm g}),
\end{eqnarray}
where $\rho_{\rm g}$ and $\rho_{\rm l}$ denote respectively the density of the gas and of the liquid. The possibility of coexistence can be visually checked by plotting in a ($\beta P, \beta \mu $) plane the high and the low density branches. A crossing of the two curves signals the 
occurrence of phase coexistence. The exact determination of the intersection point is performed numerically by using cubic spline interpolations through the calculated points.

Ideally, one should get crossings along the whole binodal line,  from lowest 
accessible temperature up to the critical point. In practice, the 
approximate nature of RHNC, shared with all integral equations, introduces
thermodynamic inconsistencies and 
modifies the singular behavior of the spinodal line obtained from 
the fluctuation route. It is transformed into a 
pseudo-spinodal line of branching points  where the real solution of 
the equations does not show a divergence of the structure factor 
$S(0)=1+\rho \tilde{h}(0)$ at zero wavenumber
or, equivalently, a vanishing of the inverse long-wavelength (lw)
isothermal compressibility
\begin{equation}
\beta \frac{\partial P}{\partial \rho} \Big |_{\rm lw} = 1-\rho \tilde{c}(0).
\end{equation}

Moreover, such a pseudo-spinodal line extends the no-solution region into a non-negligible neighborhood of the critical point, thus  preempting the possibility of finding the coexisting densities.
As a result, the possibility of getting the coexistence curve with integral equations is limited to relatively low temperatures. In a way, the quality of an integral equation can be accurately 
tested by the extent of the resulting coexistence curve.

In the case of SW fluids we have found, as discussed in the next section,  that 
reducing the width of the SW progressively extends the temperature interval below the 
critical point where the pseudo-spinodal line preempts the binodal. However, we note that the quality of the thermodynamic and structural data in the accessible  region of the phase diagram remains high. Thus, we have investigated the possibility of a small extrapolation of the RHNC data inside the pseudo-spinodal  region. In all the cases we have investigated,  such extrapolation is 
required only in the  gas phase and the resulting coexistence curve appears as a smooth continuation of the part based on  real crossings.

\begin{figure}[htbp]
\begin{center}
\includegraphics[height=9cm,angle=0]{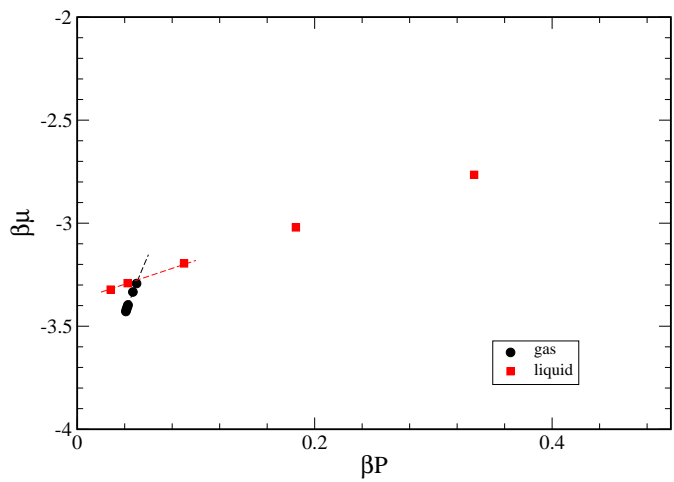}
\caption{Determination of the coexistence point. Case of quasi-crossing at $T^*=2.5$,
$\lambda=2$. Squares: liquid
branch;  circles: vapor branch. Dashed lines indicate the cubic spline
approximation used to locate the extrapolated crossing point.}
\label{fig1}
\end{center}
\end{figure}

\begin{figure}[htbp]
\begin{center}
\includegraphics[height=9cm,angle=0]{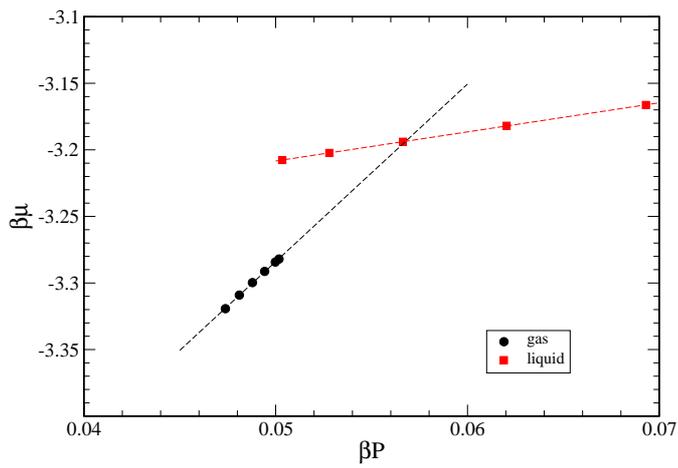}
\caption{Determination of the coexistence point. 
Case of a large extrapolation at $T^*=1.74$,
$\lambda=1.75$. Symbols as in figure 1.}
\label{fig2}
\end{center}
\end{figure}

In figure 1, we show a typical case of quasi-crossing while figure 2 shows a 
case of a more distant missed crossing. In both cases, the possibility of 
finding a coexisting vapor phase 
is preempted by the sudden disappearance of the low-density solution.
It is clear  that, although RHNC does not have a low-density solution at 
the same temperature and pressure of a corresponding liquid,  
a  smooth extrapolation of thermodynamic data could be  quite safe in cases 
like that in figure 1, 
due to the small curvature 
of the lines and to their strong transversality. Larger extrapolations, 
like that in figure
4, appear to introduce uncontrolled uncertainties. Additional comments 
will be added in the next section.

\section{Results}
As a first check of the RHNC quality, we have computed the pressure 
$P$ and excess chemical potential
\begin{equation}
\beta \mu_{\rm ex} = \frac{\beta A_{\rm ex}}{N} + \left( 
\frac{\beta P}{\rho} - 1 \right)
\end{equation}
for states at high densities with $\lambda=1.5$ examined by 
Lab\'{i}k et al. \cite{Labik} using scaled-particle Monte Carlo (SP-MC) 
simulation at $\rho^*=0.8$ and 0.9. Results for the pressure from 
simulation and the RHNC calculation are compared in figure 3, while figure 
4 gives the comparisons for the excess chemical potential. There is overall 
quite good agreement, although the weakening of the RHNC results with 
increasing density and decreasing temperature, a shortcoming common to 
all integral equation closures, does become evident.

\begin{figure}[htbp]
\begin{center}
\includegraphics[height=9cm,angle=0]{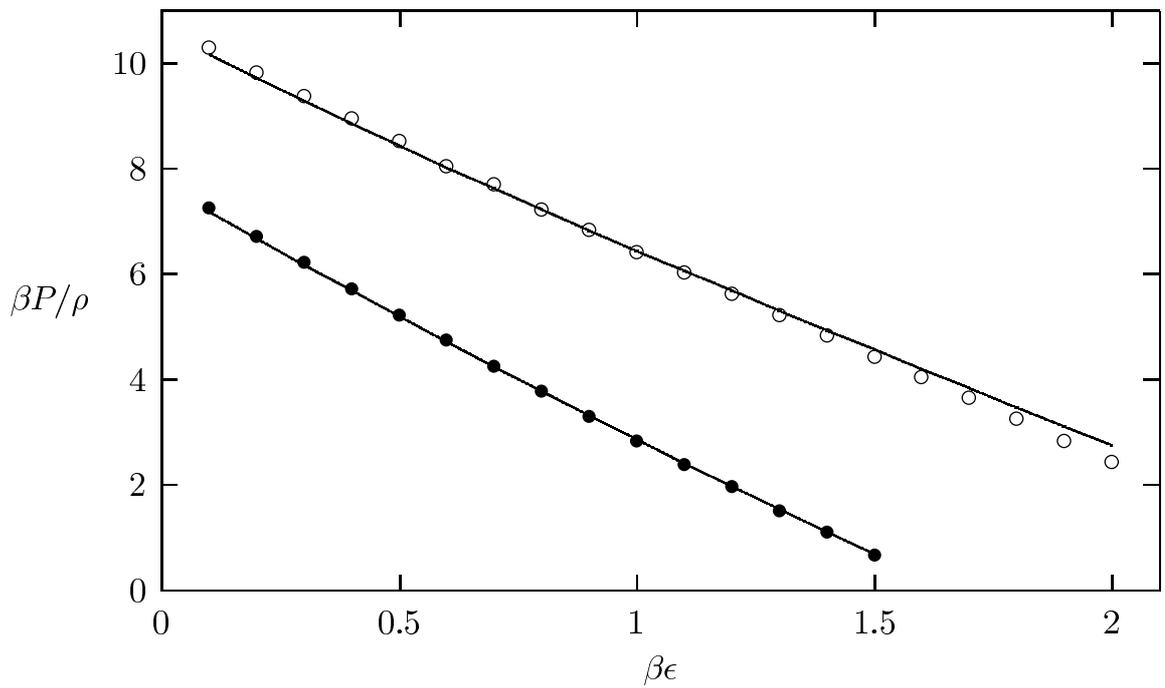}
\caption{
SP-MC pressure vs. inverse temperature for a SW fluid
with $\lambda = 1.5$ obtained by Lab\'{i}k et al. \cite{Labik}
at $\rho^*=0.8$ (filled circles) and $\rho^*=0.9$
(empty circles). Solid lines are the RHNC results.
}
\label{fig3}
\end{center}
\end{figure}%

\begin{figure}[htbp]
\begin{center}
\includegraphics[height=9cm,angle=0]{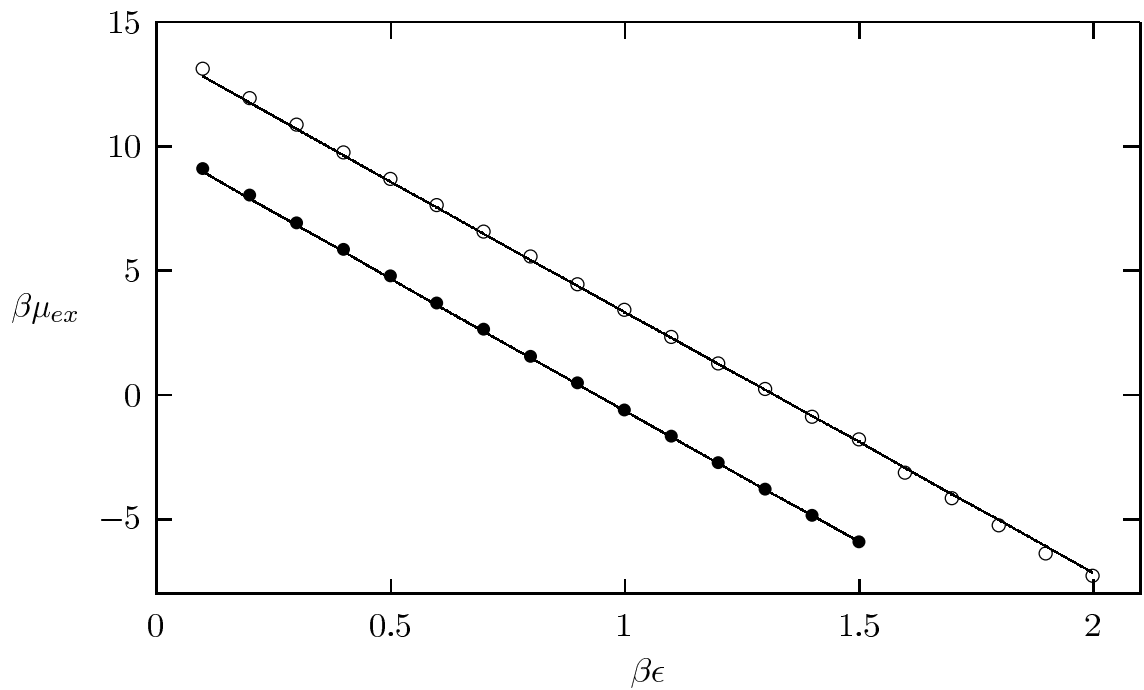}
\caption{
SP-MC chemical potential vs. inverse temperature for a SW fluid
with $\lambda = 1.5$ obtained by Lab\'{i}k et al. \cite{Labik}
at $\rho^*=0.8$ (filled circles) and $\rho^*=0.9$
(empty circles). Solid lines are the RHNC results.
}
\label{fig4}
\end{center}
\end{figure}%

In tables 1 and 2  we report a few significant comparisons between 
RHNC  thermodynamic and structural 
results and recent numerical simulation data \cite{Largo2,Largo} for SW 
systems of different range.

\begin{table}
\caption{Comparison of simulation and RHNC thermodynamic quantities of square-well fluids. For each density, the first row contains Monte Carlo data of Largo 
and Solana \cite{Largo2} and the second row our results.}

\centering
\begin{tabular}{l c c c c }
\hline 
\hline 
& & & & \\
& & \mbox{$\beta P/\rho$ \quad-$U/N\epsilon$} &\mbox{$\beta P/\rho$ \quad -$U/N\epsilon$}  &\mbox{$\beta P/\rho$ \quad -$U/N\epsilon$} \\
& & & & \\
$\lambda=1.2$ & $\rho^*$ & $T^*=0.7$ & $T^*=1.0$ & $T^*=3.0$  \\
& & & & \\
~ & 0.1 & \mbox{0.76 \quad 0.62} & \mbox{0.96 \quad 0.40} &\mbox{1.17 \quad 0.23}  \\
~ &     & \mbox{0.76 \quad 0.62} & \mbox{0.96 \quad 0.40} &\mbox{1.17 \quad 0.23}  \\
& & & & \\
~ & 0.8 & \mbox{0.77 \quad 3.68} & \mbox{2.72 \quad 3.43} &\mbox{6.03 \quad 3.09}  \\
~ &     & \mbox{0.68 \quad 3.58} & \mbox{2.68 \quad 3.38} &\mbox{5.99 \quad 3.09}  \\
& & & & \\
$\lambda=1.5$ &   & $T^*=1.5$ & $T^*=3.0$ &  \\
& & & & \\
~ & 0.1 & \mbox{0.78 \quad 0.92} & \mbox{1.03 \quad 0.69} &  \\
~ &     & \mbox{0.78 \quad 0.92} & \mbox{1.03 \quad 0.69} &  \\
& & & & \\
~ & 0.7 & \mbox{2.31 \quad 5.28} & \mbox{4.02 \quad 5.08} &  \\
~ &     & \mbox{2.30 \quad 5.27} & \mbox{3.98 \quad 5.07} &  \\
& & & & \\
$ \lambda=2.0 $ & & $T^*=3.0$ & $T^*=5.0$ &  \\
& & & & \\
~ & 0.1 & \mbox{0.69 \quad 1.96} & \mbox{0.91 \quad 1.72} &  \\
~ &     & \mbox{0.69 \quad 1.98} & \mbox{0.92 \quad 1.73} &  \\
& & & & \\
~ & 0.6 & \mbox{1.19 \quad 9.75} & \mbox{2.44 \quad 9.58} &  \\
~ &     & \mbox{1.18 \quad 9.76} & \mbox{2.43 \quad 9.59} &  \\
& & & & \\
~ & 0.7 & \mbox{2.01 \quad 11.35} & \mbox{3.51 \quad 11.16} &  \\
~ &     & \mbox{1.97 \quad 11.35} & \mbox{3.47 \quad 11.16} &  \\
& & & & \\
\hline 
\hline 
\end{tabular}
\end{table}

\begin{table}
\caption{ 
Comparison of simulation and RHNC contact values $g(\sigma^+)$,
$g(\lambda\sigma^-)$ and $g(\lambda\sigma^+)$  of the radial distribution 
function of square-well 
fluids. Monte Carlo data (MC) from Largo et al. \cite{Largo}.
}

\centering
\begin{tabular}{l c c c c c c c c}
\hline
\hline
& & & & & & & &\\
& & & \multicolumn{2}{c}{$g(\sigma^+)$} & \multicolumn{2}{c}{$g(\lambda\sigma^-)$} & \multicolumn{2}{c}{$g(\lambda\sigma^+)$} \\
$\lambda$ & $T^*$ & $\rho^*$ & MC & RHNC  & MC & RHNC & MC & RHNC  \\
\hline
1.1 & 0.5 & 0.1 & 7.254 & 7.260 & 7.179 & 7.179 & 0.966 & 0.973 \\
    &     & 0.5 & 6.088 & 5.835 & 5.816 & 5.460 & 0.787 & 0.739 \\
    &     & 0.7 & 5.667 & 5.681 & 5.307 & 4.967 & 0.718 & 0.672 \\
1.2 & 0.7 & 0.1 & 4.126 & 4.150 & 4.030 & 4.046 & 0.968 & 0.970 \\
    &     & 0.5 & 3.449 & 3.272 & 3.122 & 2.945 & 0.748 & 0.706 \\
    &     & 0.7 & 3.281 & 3.186 & 2.810 & 2.744 & 0.675 & 0.658 \\
1.5 & 1.5 & 0.1 & 1.952 & 1.957 & 1.832 & 1.835 & 0.941 & 0.944 \\
    &     & 0.5 & 1.989 & 1.994 & 1.382 & 1.382 & 0.709 & 0.709 \\
    &     & 0.7 & 2.783 & 2.770 & 1.115 & 1.149 & 0.590 & 0.590 \\
    & 2.0 & 0.1 & 1.661 & 1.655 & 1.563 & 1.561 & 0.945 & 0.947 \\
    &     & 0.5 & 2.006 & 2.003 & 1.274 & 1.272 & 0.771 & 0.772 \\
    &     & 0.7 & 2.880 & 2.865 & 1.063 & 1.061 & 0.646 & 0.644 \\
1.8 & 2.0 & 0.1 & 1.839 & 1.914 & 1.599 & 1.642 & 0.974 & 0.996 \\
    &     & 0.5 & 2.199 & 2.211 & 1.175 & 1.181 & 0.714 & 0.716 \\
    &     & 0.7 & 3.487 & 3.474 & 1.141 & 1.147 & 0.693 & 0.695 \\
    & 3.0 & 0.1 & 1.449 & 1.456 & 1.320 & 1.320 & 0.942 & 0.945 \\
    &     & 0.5 & 2.162 & 2.162 & 1.079 & 1.080 & 0.773 & 0.774 \\
    &     & 0.7 & 3.392 & 3.365 & 1.043 & 1.045 & 0.747 & 0.751 \\
& & & & & & & &\\
\hline
\hline
\end{tabular}
\end{table}

We notice a progressive worsening of thermodynamic 
results for stronger couplings, 
meaning lower temperatures and higher densities, and for 
decreasing $\lambda$. Such behavior as a function 
of well width is understandable, since in the limit of very large 
$\lambda$ 
and vanishing $\epsilon$, 
the properties of SW fluid should be well described by HS physics, 
supplemented by the 
van der Waals mean field approximation. 
In such a limit RHNC should provide very accurate results.
In the opposite case, we would expect the maximum deviation from the 
hypothesis of universality of the bridge function and then the maximum 
discrepancy from simulation. In the only previous RHNC study of SW fluid 
that we are aware of, ref. \cite{Gil-Villegas}, a limited analysis of the 
differences between HS and SW  bridge functions has been attempted. 
In that paper, the authors assumed the same functional form for the bridge 
function outside the hard core  for both HS and SW systems and determined 
the parameters via a mean-square fitting of Monte Carlo (MC) data via RHNC. 
Apparently, optimal  SW bridge functions should be  slightly smaller than 
the best HS functions for the same system, but we note that the 
constrained procedure did not allow for qualitative changes of the 
functional form. We can add that the most important  for possible changes is 
the region within and close to the hard core. Direct evidence for this 
comes from a few calculations we performed by modifying the long range 
part of the bridge function in a very crude way:  we made it vanish starting 
from its first zero. The resulting RHNC  solution was almost unaffected 
by this change, thus confirming that any future effort to improve the 
bridge function beyond the HS approximation should be focused on the core 
interval between zero and the HS diameter. 

\begin{figure}[htbp]
\begin{center}
\includegraphics[height=9cm,angle=0]{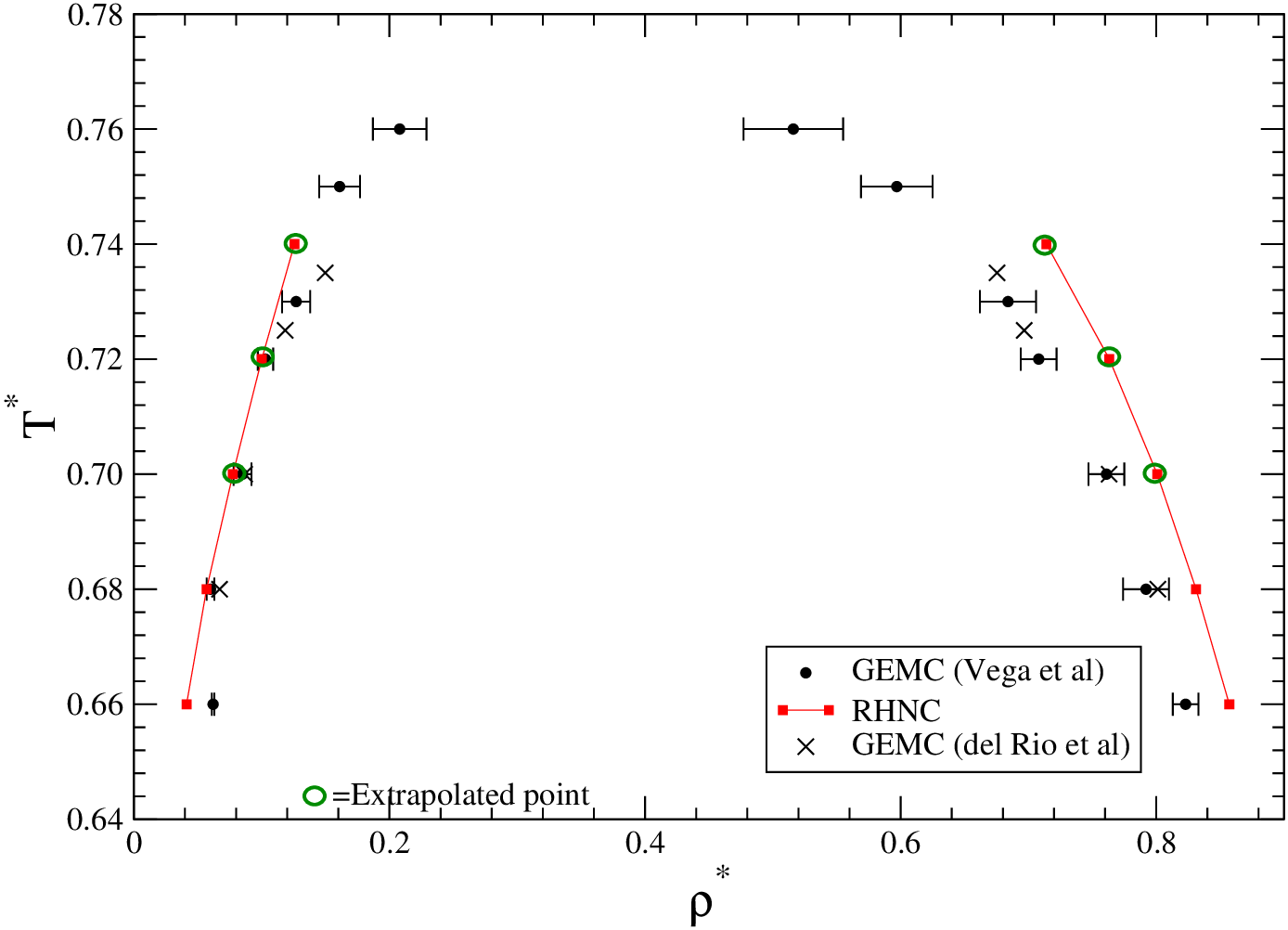}
\caption{Vapor-liquid $T^*$ versus $\rho^*$ coexistence for the SW fluid of range
$\lambda=1.25$: squares, this work (circled squares indicate extrapolation 
of the vapor branch), a line has been drawn through the points as a visual 
guide; dots, Vega et al. \cite{Vega}; crosses, 
del R\'{\i}o et al. \cite{delRio}.}
\label{fig5}
\end{center}
\end{figure}
\begin{figure}[htbp]
\begin{center}
\includegraphics[height=9cm,angle=0]{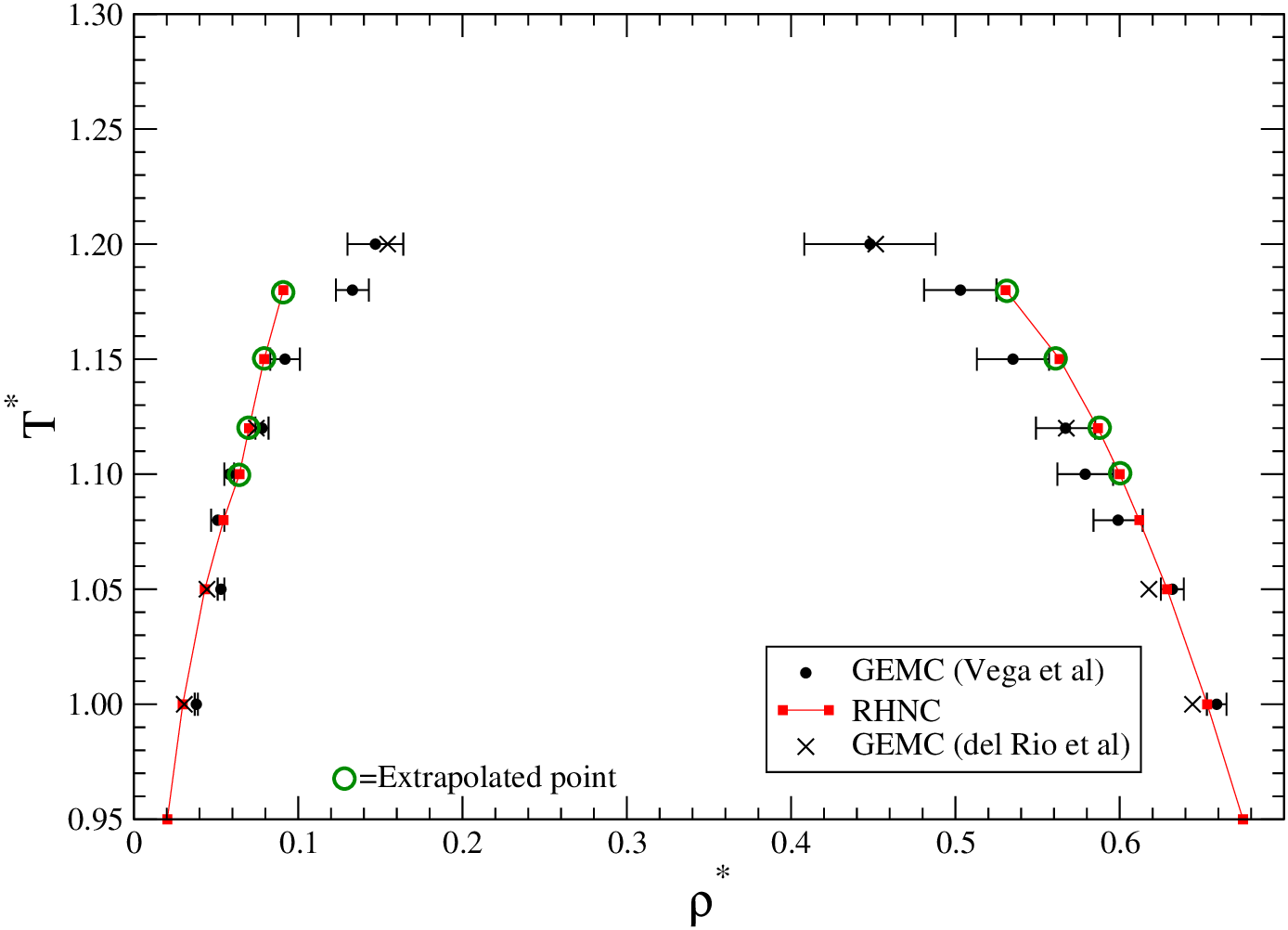}
\caption{Vapor-liquid $T^*$ versus $\rho^*$ coexistence for the SW fluid of range
$\lambda=1.5$: squares, this work (circled squares indicate extrapolation of the vapor branch), a line has been drawn through the points as a visualguide; dots, Vega et al. \cite{Vega}; crosses, del R\'{\i}o et al. \cite{delRio}.}
\label{fig6}
\end{center}
\end{figure}
\begin{figure}[htbp]
\begin{center}
\includegraphics[height=9cm,angle=0]{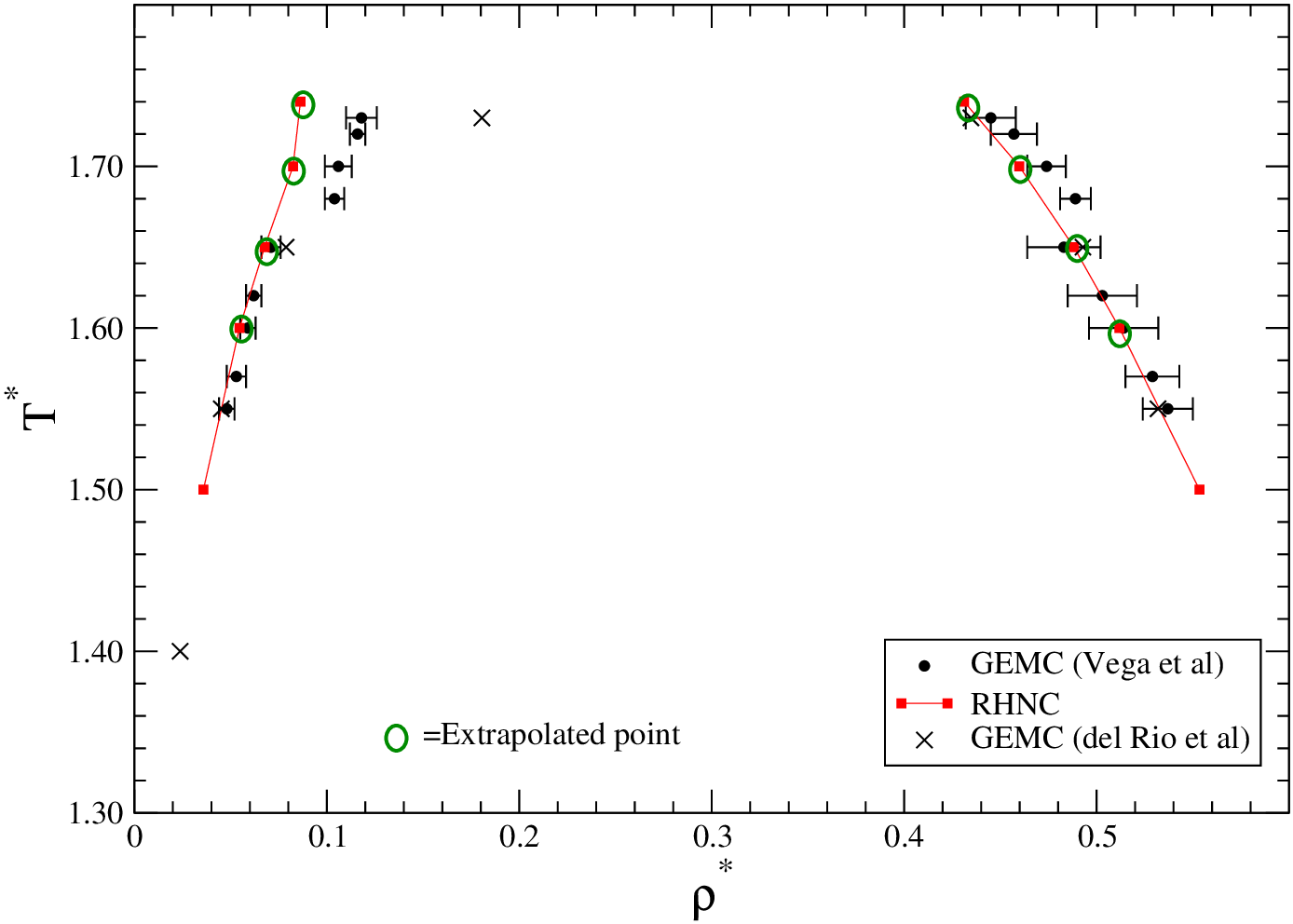}
\caption{Vapor-liquid $T^*$ versus $\rho^*$ coexistence for the SW fluid of range
$\lambda=1.75$: squares, this work (circled squares indicate extrapolation of the vapor branch), a line has been drawn through the points as a visualguide); 
dots, Vega et al. \cite{Vega}; crosses, del R\'{\i}o et al. \cite{delRio}.}
\label{fig7}
\end{center}
\end{figure}
\begin{figure}[htbp]
\begin{center}
\includegraphics[height=9cm,angle=0]{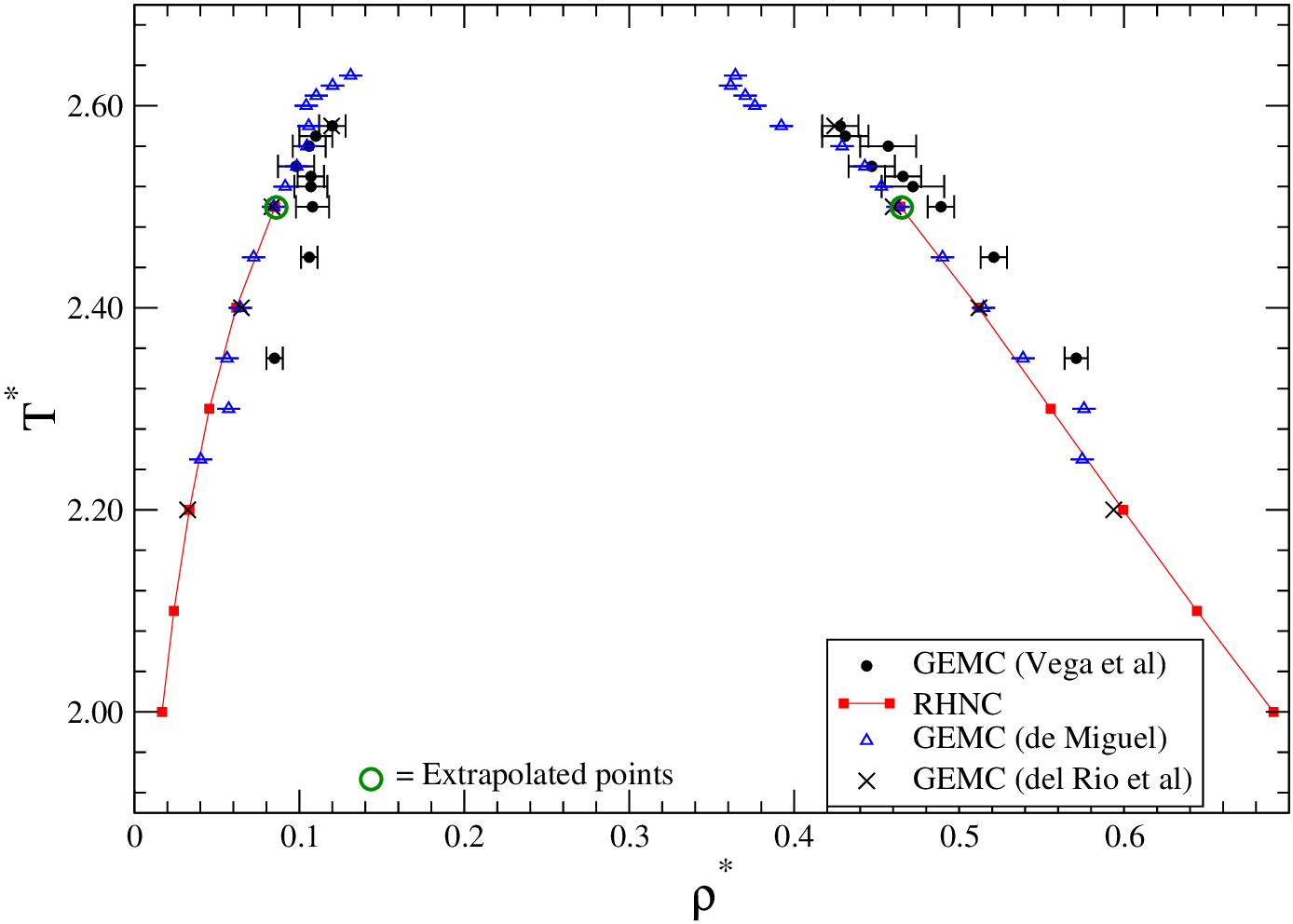}
\caption{Vapor-liquid $T^*$ versus $\rho^*$ coexistence for the SW fluid of range
$\lambda=2.0$: squares, this work (circled squares indicate extrapolation of the vapor branch), a line has been drawn through the points as a visualguide; dots, Vega et al. \cite{Vega}; crosses, del R\'{\i}o et al. \cite{delRio}; triangles, de Miguel \cite{deMiguel}.}
\label{fig8}
\end{center}
\end{figure}
We have studied the RHNC liquid-vapor coexistence of four SW systems of 
width size $\lambda =$ 1.25, 1.5, 1.75, and 2. In figures 5-8  we present 
our results for the coexistence curves of the four SW fluids we have 
examined. RHNC data (squares connected by a line) are compared with 
different sets of 
simulation data (see figure captions). All the cases where RHNC did 
not provide a real crossing between the two branches (low and high density) 
of the $\beta P$-$\beta \mu$ curves have been marked by a circle around the 
square. 

Two main features of the results are evident: the overall semiquantitative 
accuracy of the coexistence curve in all the investigated cases and the 
progressive worsening of the quality of the RHNC results as the range 
$\lambda$ is reduced. The latter appears as the obvious consequence, 
at the level of a phase diagram, of our observations about the quality of
thermodynamic data as a function of the range. At $\lambda=2$,  our results 
closely follow the data of de Miguel \cite{deMiguel} and of del 
R\'{\i}o et al. \cite{delRio}, while Gibbs ensemble Monte Carlo results by 
Vega at al. \cite{Vega} seem to be definitely biased toward higher densities. In addition, the unphysical changes of the curvature seen in the MC data 
suggest that one might consider quite optimistic the published statistical 
error bars.

At $\lambda=1.75$, RHNC results start to show clear evidence of a distortion 
of the coexistence curve at the highest temperature,  although all the points 
below $T^*=1.70$ are in good agreement with MC data. Since the  extrapolation 
at this temperature corresponds to the case shown in figure 2, while the 
remaining extrapolations correspond to closer missed intersections, we have 
a measure of the quantitative effect of our extrapolation procedure on the 
resulting binodal line.

Data at $\lambda=1.5$ and $\lambda=1.25$ clearly show a progressive 
decrease of the quality on the high density side of the curve. As previously discussed, 
this region corresponds to the region where RHNC could be substantially 
affected by improvements in the description of the bridge function.
We stress that coexistence is a severe test for the quality of integral equations. In particular, pressure and chemical potential are quantities quite sensitive to small changes of the closure.

However, according to Liu et al. \cite{Liu}, at $\lambda=1.25$ the 
liquid-vapor transition becomes metastable and is preempted by the 
freezing transition. Thus, the observed decreased quality at smaller widths 
is somewhat compensated by the fact that the integral equation 
results for such case are referring to a metastable liquid.

In all the cases, integral equations do not allow one to get close enough to 
the critical point to provide a direct estimate of its location. 
However, a good quality in reproducing the low temperature part of the liquid 
vapor coexistence and the hypothesis of Ising-like value for the critical 
exponent ($\beta = 0.325$) may provide a first approximation for the extent 
of the two-phase region.

\section*{Conclusions}

We have shown that RHNC is able to reproduce the low-temperature part of the liquid-vapor phase diagrams of square-well fluids. With increasing temperature and depending on the extent of the attractive well, there is a region of the binodal line that can be obtained  by a slight extrapolation of the vapor phase beyond the numerical limit of the quasi-spinodal line. Eventually, approaching the critical temperature, the extrapolation becomes very poor and the shape of the coexistence line is strongly affected. The agreement with the best computer simulations is good for the  widest well sizes, but becomes less satisfactory for $\lambda$ at or below 1.5. Since the only approximation of the theory is the bridge function (apart from inherent limitations of numerical computation), our results point to the need for an improvement of the short range part of the 
bridge function.

It is interesting to note also that our results are consistent with 
similar RHNC  calculations on the Lennard-Jones fluid \cite{Caccamo,Lomba}, 
where it is possible to obtain
a larger part of the binodal line than in the case of $\lambda=2$, 
thus confirming our findings on better performances for wider range potentials.

We stress again that the main purpose of the present investigation is not 
to propose RHNC as an alternative to more refined theories such as SCOZA or HRT
when a highly accurate description of the critical properties is required. 
Rather, we have the more limited goal of checking the ability of the RHNC formulation to 
provide an approximate but quick tool for locating the liquid-vapor phase 
transition in SW-like systems. In all the isotropic SW cases we have studied, the theory allows one to predict from scratch an unknown phase coexistence in a computational time that is negligible compared to numerical simulations, using existing numerical techniques for integral equation solutions. Such a feature may only be marginally interesting for isotropic SW fluids, but becomes crucial for anisotropic patchy models of globular macromolecules. We are currently investigating this topic.

After this paper was submitted for publication, a very thorough work on the 
same topic of liquid-vapor phase equilibrium for the SW fluid using a 
different integral equation was published by El Mendoub et al. \cite{Mendoub}. 
These authors use an efficient ``adaptive grid'' \cite{Charpentier} 
that automates the mapping of the no-solution space in the temperature-density 
plane. Comparing their results to ours, we notice that RHNC results for the
radial distribution function show a uniform better agreement with simulation 
data. Since deviations from computer simulation binodal lines shown by their closure 
and the present study have  opposite directions, a detailed comparison between their closure and RHNC would be interesting . 

\bigskip
{\bf Acknowledgments}

\smallskip
A.G. acknowledges financial support from Miur-Cofin 2008/2009.

\end{document}